\documentclass[aps,prd,preprint,groupedaddress,nofootinbib,showpacs]{revtex4}

\def\lb{\lbrack}
\def\rb{\rbrack}

 \def\Slash#1{
  \begin{picture}(5,6)(0,0)
  \put(-.7,-1.2){\line(5,6)6}
  \end{picture}
  \kern-.8em#1}
 \def\slash#1{
  \begin{picture}(5,6)(0,0)
  \put(-1.5,-1.7){\line(5,6)5}
  \end{picture}
  \kern-.8em#1}

\def\gg5{\gamma_5}
\def\hg5{\hat{\gamma}_5}
\def\g4{\gamma_4}

\def\F{{\cal F}}

\def\D{{\cal D}}
\def\P{{\cal P}}
\def\F{{\cal F}}

\def\wR{\widetilde{R}}

\def\Qlatmr1{Q_{lat}^{(m=r=1)}}

\def\be{\begin{eqnarray}}
\def\ee{\end{eqnarray}}

\def\hA{\hat{A}}
\def\hmu{\hat{\mu}}

\def\gmu{\gamma_{\mu}}
\def\gnu{\gamma_{\nu}}

\def\pmu{\partial_{\mu}}
\def\pnu{\partial_{\nu}}

\def\bMS{\overline{MS}}

\begin{document}

\input epsf

\title{Relation between bare lattice coupling and $\bMS$ coupling
at one loop with general lattice fermions}

\author{David H.~Adams}
\email{dadams@phya.snu.ac.kr}

\affiliation{Dept. of Physics and Astronomy, Seoul National University, 
Seoul, 151-747, South Korea}

\altaffiliation{Current address}

\affiliation{Physics Dept., National Taiwan University, 
Taipei 106, Taiwan R.O.C.}

\affiliation{Physics Dept., Florida International University, 
11200 SW 8th St., Miami, FL 33199, U.S.A.}

\date{April 26, 2008}

\begin{abstract}

A compact general integral formula is derived from which the fermionic contribution to the 
one-loop coefficient in the perturbative expansion of the $\bMS$ coupling in powers
of the bare lattice coupling can be extracted. It is seen to reproduce the known results
for unimproved naive, staggered and Wilson fermions, and has advantageous features which
facilitate the evaluation in the case of improved lattice fermion formulations. This is
illustrated in the case of Wilson clover fermions, and an expression in terms of known lattice 
integrals is obtained in this case which gives the coefficient to much greater
numerical accuracy than in the previous literature.

\end{abstract}

\pacs{11.15.Ha, 11.25.Db}

\maketitle

\section{Introduction}

When transforming results from lattice simulations into a continuum scheme such as $\bMS$
it is often desirable to know the perturbative expansion of the renormalized coupling 
in powers of the bare lattice coupling. This is useful as an intermediate step for relating
the $\bMS$ coupling to the coupling defined in nonperturbative lattice schemes such as
the ones based on the static quark potential \cite{Lepage-Mackenzie,Trottier(review)}
and Schr\"odinger functional \cite{L-W(Schrodinger),Bode-Weisz}, and is also needed to 
translate bare lattice quark masses into the $\bMS$ scheme (see, e.g., 
\cite{Sachrajda,staggeredmass}). 
The one loop coefficient in the expansion is of further interest because it determines the ratio 
of the lattice and $\bMS$ $\Lambda$-parameters 
\cite{Has(L)(PLB),DashenGross,Has(L)(NPB),Kawai,Weisz(PLB),Weisz(NPB),Vicari,Skour-Pan}.
Moreover, the one loop coefficient
is also needed for determining the two loop relation between the couplings,
from which the third term in the lattice beta-function (governing the approach to the continuum 
limit) can be determined \cite{L-W(Schrodinger),Alles,Christou,Bode-Pan,Const-Pan1,Const-Pan2}.

In this paper we derive, for general lattice fermion formulation, a compact general integral 
formula from which the fermionic contribution to the one-loop coefficient in the perturbative 
expansion of the $\bMS$ coupling in powers of the bare lattice coupling can be extracted.
The motivations for pursuing this are as follows. First, 
given the plethora of lattice fermion actions currently in use, and the likelyhood of 
new ones or improved versions of present ones being developed in the future, it is desirable
where possible to have general formulae from which quantities of interest can be calculated 
without having to do the calculation from scratch each time.
Second, it is desirable to have independent ways to check the computer programs used
these days to perform lattice perturbation theory calculations via symbolic manipulations.
Third, by reducing the calculation to a managable number of
one loop lattice integrals one can more easily achieve greater numerical precision than
with symbolic computer programs. 
This is important, since, as emphasized in \cite{Capitani}, the one loop results need to be 
determined with very high precision to achieve reasonable precision in the two loop result.
As a demonstration that the general formulae of this paper are useful in this regard,
we apply them to obtain the fermionic contribution to the one loop coefficient in the case of 
Wilson clover fermions \cite{clover} to almost twice as many significant decimal places as in the 
previous literature.



\begin{figure}

$$
\epsfysize=6cm \epsfbox{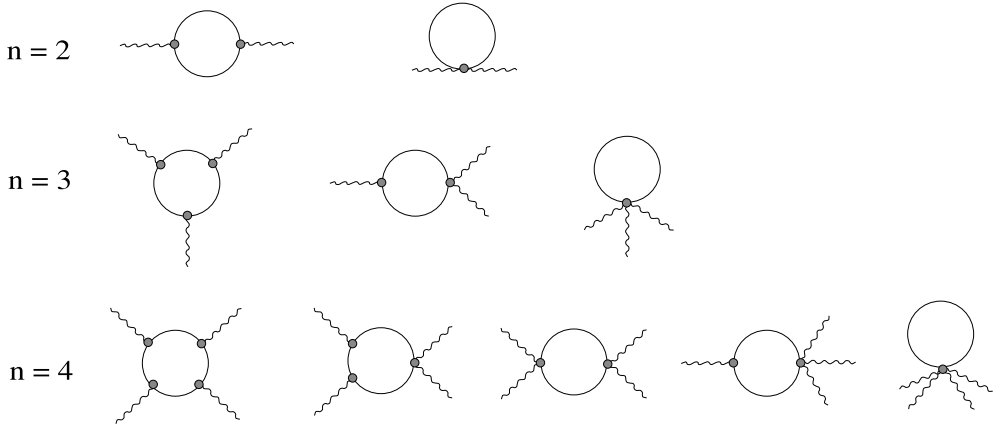}
$$
\caption{The lattice Feynman diagrams for the gluonic $n$-point function with 
internal fermion loop for $n\!=\!2,3,4$.}
\label{vac1}
\end{figure}

As reviewed in sect.~II,
determining the fermionic contribution to the one loop coefficient reduces to determining a
constant $c_I$ arising in a logarithmically divergent one fermion loop lattice Feynman integral
$I(am)$, which has the general structure
\be
I(am)=\frac{1}{24\pi^2}\log(a^2m^2)+c_I
\label{1.1}
\ee
Here $a$ is the lattice spacing and $m$ an infrared regulator fermion mass. The numerical factor 
in the log term is universal, whereas $c_I$ depends on the details of the lattice fermion
formulation. $I(am)$ arises from the one fermion loop contribution to the gluonic 2-point 
function, and it is from this that it was evaluated in previous works for specific 
lattice fermion formulations. 
However, Ward Identities allow $I(am)$ to also be evaluated from the gluonic 3- or
4-point functions. In this paper we evaluate $I(am)$ from the one fermion loop contribution
to the gluonic 4-point function. In this case there are five lattice Feynman diagrams to consider
rather than the two diagrams for the gluonic 2-point function -- see Fig.~\ref{vac1}.
Nevertheless, evaluation of $I(am)$ from the 4-point function turns out to be advantageous.
The diagrams are evaluated at vanishing external momenta without the need to first take 
momentum derivatives, and we find three nice properties: \hfill\break
(i) Only one of the five diagrams is logarithmically divergent -- it is the first $n\!=\!4$ 
diagram in Fig.~\ref{vac1}. The other four diagrams are all convergent. \hfill\break
(ii) The logarithmically divergent diagram is not affected by changes in how the 
link variables are coupled to the fermions (e.g., it is unchanged by adding staples, clover term
etc.). Consequently, it is the same for improved and unimproved versions of the lattice fermion 
formulation (provided the free field formulations are the same). \hfill\break
(iii) The four convergent diagrams, or subsets of them, vanish when the lattice Dirac operator 
is sufficiently simple. In particular, they all vanish for unimproved Wilson and 
staggered fermions, also when the Naik term \cite{Naik} is included. \hfill\break 
Thus for improved versions of Wilson and staggered fermions the only new quantities to compute 
relative to the unimproved case are the four convergent one-loop lattice 
integrals.\footnote{This is only true in the staggered fermion case when the Naik term is not
used. More on this in the concluding section.}

The main result in this paper is a general integral formula for $I(am)$ obtained by evaluating
the contributions from the five $n\!=\!4$ Feynman diagrams in Fig.~\ref{vac1} for general lattice
fermion formulation, from which the desired constant $c_I$ can be extracted.
Specifically, we do the following: (a) evaluate the contribution from the logarithmically 
divergent diagram, deriving a quite explicit general formula which is seen to reproduce previous
results for the cases of unimproved Wilson and naive/staggered fermions, and (b) derive formulae
for, and describe a straightforward procedure for evaluating, the contributions from the four 
convergent diagrams. We illustrate this in the case of Wilson clover fermions.
The general formulae lead to integrals to which
the method of Ref.~\cite{Burgio} can be applied, reducing the integrals to basic lattice integrals 
that are already known to high precision.
The application of our result to other lattice fermion formulations such as Asqtad staggered 
fermions \cite{asqtad} and overlap fermions \cite{overlap} will be made in future work.

The paper is organized as follows. Sect.~II reviews the one loop expansion of the $\bMS$ 
coupling in the bare lattice coupling, using the background field approach. 
In sect.~III we derive an initial expression for $I(am)$ as the sum of contributions from the five 
$n\!=\!4$ diagrams in Fig.~1 and point out the properties (i), (ii) and (iii) mentioned above. 
Rather than evaluating
the diagrams directly, we infer them from perturbative expansion of the fermion determinant,
which is easier. From this the general formulae and applications mentioned in (a) and (b) above
are derived in sect.~IV and V, respectively.
The concluding sect.~VI describes applications to be carried out in future work, as well as
the possibility of deriving similar results for the gauge--ghost contribution to the one loop
coefficient. Also, it is pointed out that the present results are relevant for a previous proposal 
for constructing the gauge field action on the lattice from the lattice fermion determinant 
\cite{Horvath}. Specifically, our results give an expression for the coefficient of the Yang-Mills
action that arises in that proposal. Some technical details of our calculations are provided
in two appendices.

\section{Generalities of the one-loop relation betwen bare lattice coupling and $\bMS$
coupling}

The gauge field quantum effective action $\Gamma(A)$ in Euclidean spacetime can be expressed 
(prior to gauge fixing) as
\be
e^{-\Gamma(A)}=\int\D Q\D\bar{\psi}\D\psi\,
e^{-\frac{1}{g^2}S_{YM}(A+gQ)-\int\bar{\psi}D(A+gQ)\psi} 
\label{2.1}
\ee
where $S_{YM}$ is the Yang-Mills action and $D(A+gQ)$ is the Dirac operator coupled to 
$A+gQ$ where $A$ is the background gauge field and $Q$ the quantum fluctuation field.
The gauge group is taken to be SU(N) with the fermion fields being in the fundamental 
representation. 
$\Gamma(A)$ has the loop expansion
\be
\Gamma(A)=\frac{1}{g^2}S_{YM}(A)+\Gamma_1(A)+g^2\Gamma_2(A)+\dots+g^{2(n-1)}\Gamma_n(A)+\dots
\label{2.2}
\ee
Expanding $\Gamma(A)$ in powers of $A$, the term quadratic in $A$ has the form
\be
\Gamma^{(2)}(A)=\int_p\Gamma^{(2)}(p)_{\mu\nu}^{ab}\,\hA_{\mu}^a(p)\hA_{\nu}^b(-p)
\label{2.3}
\ee
where $\hA_{\mu}^a(p)$ denotes the Fourier transform of $A_{\mu}^a(x)$. Gauge invariance of 
$\Gamma(A)$ and rotation symmetry imply that the 2-point function has the form
\be
\Gamma^{(2)}(p)_{\mu\nu}^{ab}=-\delta^{ab}(p^2\delta_{\mu\nu}-p_{\mu}p_{\nu})
\Big\lb\frac{1}{g^2}-\nu_1(p)-\dots-g^{2(n-1)}\nu_n(p)-\dots\Big\rb
\label{2.4}
\ee

The relation between the $\bMS$ renormalized coupling $g$ and bare lattice coupling 
$g_0$ at one loop can be obtained by requiring equality of $\Gamma(A)_{\bMS}$ and 
$\Gamma(A)_{lat}$ up to one loop \cite{DashenGross,Has(L)(NPB)}:
\be
\frac{1}{g^2}S_{YM}(A)+\Gamma_1(A)_{\bMS}=\frac{1}{g_0^2}S_{YM}(A)+\Gamma_1(A)_{lat}  
\label{2.5}
\ee
Expanding each side in powers of $A$, the equality betwen the quadratic terms gives, in light
of (\ref{2.4}),\footnote{In the lattice theory there is only hypercubic rotation symmetry.
However, this suffices to obtain (\ref{2.4}) in the lattice setting up to terms which vanish
for $a\to0$.}
\be
\frac{1}{g^2}-\nu_1(p)_{\bMS}=\frac{1}{g_0^2}-\nu_1(p)_{lat}
\label{2.6}
\ee
The one-loop contribution to the 2-point function is given by a sum of 3 terms: gauge field loop,
ghost loop and fermion loop. Consequently, for $N_f$ flavors of massless fermions, 
$\nu_1(p)_{\bMS}$ and $\nu_1(p)_{lat}$ have the general forms
\be
\nu_1(p)_{\bMS}&=&-\beta_0\log(p^2/\mu^2)+c_{\bMS}^g+N_f\,c_{\bMS}^f
\label{2.7} \\
\nu_1(p)_{lat}&=&-\beta_0\log(a^2p^2)+c_{lat}^g+N_f\,c_{lat}^f
\label{2.8} 
\ee
where $\mu$ is the mass scale in the $\bMS$ scheme, $a$ is the lattice spacing, and
\be
\beta_0=\frac{1}{16\pi^2}\Big(N\frac{11}{3}-N_f\frac{2}{3}\Big)
\label{2.9}
\ee
The constant $c^g$ arises from the gauge and ghost loop contributions; it depends on $N$ and
the gauge fixing parameter, while $c^f$ arises from the fermion loop contribution. In the lattice
case $c^g$ and $c^f$ also depend on whatever parameters are present in the lattice gauge and
fermion actions. (E.g., for lattice Wilson fermions $c_{lat}^f$ depends on the Wilson
parameter.)

From (\ref{2.6}) and (\ref{2.7})--(\ref{2.8}) we get
\be
\frac{1}{g^2}=\frac{1}{g_0^2}+\beta_0\log(a^2\mu^2)+l_0
\label{2.10}
\ee
where 
\be
l_0&=&l_0^g+N_fl_0^f \label{2.11} \\
l_0^g=c_{\bMS}^g-c_{lat}^g\ &,&\qquad l_0^f=c_{\bMS}^f-c_{lat}^f
\label{2.12}
\ee
(Our notations $l_0\,$, $l_0^g\,$, $l_0^f$ follow the papers of Panagopoulos and collaborators,
e.g., \cite{Skour-Pan})
The relation bewteen $g$ and $g_0$ up to one loop now follows from (\ref{2.10}):
\be
g^2=g_0^2(1-g_0^2\lb\beta_0\log(a^2\mu^2)+l_0\rb+O(g_0^4)\,)
\label{2.13}
\ee
Also, from (\ref{2.10}) or (\ref{2.13}) the relation between the lattice and $\bMS$ 
$\Lambda$-parameters is obtained 
\cite{Has(L)(PLB),DashenGross,Has(L)(NPB),Kawai,Weisz(PLB)}:
\be
\frac{\Lambda_{lat}}{\Lambda_{\bMS}}=e^{l_0/2\beta_0}
\label{2.14}
\ee

The focus of our attention in this paper is the fermionic contribution to $l_0\,$,
i.e., $l_0^f$ in (\ref{2.12}) for general lattice fermion formulation. 
The continuum constant $c_{\bMS}^f$ is well-known (see, e.g., \cite{Christou}):
\be
c_{\bMS}^f=-\frac{5}{72\pi^2}
\label{2.15}
\ee
Therefore, to determine $l_0^f$ we need to determine the lattice constant $c_{lat}^f$ for 
general lattice fermion formulation.

To study $c_{lat}^f$ it suffices to consider the $N_f=1$ case which we restrict to henceforth.
We proceed by expanding the lattice 2-point function $\Gamma_{lat}^{(2)}(p)_{\mu\nu}^{ab}$ 
in powers of (the components of) the external momentum $p$. Although 
$\Gamma_{lat}^{(2)}(p)_{\mu\nu}^{ab}$ itself is infrared finite, the terms in the expansion
are individually infrared divergent. To deal with this we introduce a fermion mass $m$ as
regulator; it renders the expansion terms finite.
Since gauge invariance is maintained when $m$ is introduced, the expansion of the one fermion
loop contribution to $\Gamma_{lat}^{(2)}(p)_{\mu\nu}^{ab}$ up to 2nd order in $p$ results
in an expression of the general form
\be
\Big\lb\Gamma_{lat}^{(2)}(p)_{\mu\nu}^{ab}\Big\rb_{f,\,1-loop}
=\delta^{ab}(p^2\delta_{\mu\nu}-p_{\mu}p_{\nu})\lb I(am)+R(p^2/m^2)\rb
\label{2.16}
\ee
up to terms which vanish for lattice spacing $a\to0$. Here $I(am)$ is a logarithmically 
divergent one loop lattice Feynman integral given by
\be
I(am)=-\partial_{\mu}\partial_{\nu}\Big\lb\Gamma_{lat}^{(2)}(p)_{\mu\nu}^{11}
\Big\rb_{f,\,1-loop}\Big|_{p=0}\quad\quad\mbox{($\mu\ne\nu\,,\ $ no sum)}
\label{2.17}
\ee
($\partial_{\mu}\equiv\frac{\partial}{\partial p_{\mu}}$). By the structural  
result of \cite{DA(logdiv)} it has the general form\footnote{The factor $1/24\pi^2$ in 
the log term is fixed by universality \cite{DA(logdiv)}; it is minus the fermionic term in 
$\beta_0$ with $N_f\!=\!1$.}
\be
I(am)=\frac{1}{24\pi^2}\log(a^2m^2)+c_I
\label{2.18}
\ee
where the constant $c_I$ depends on the details of the lattice fermion formulation.
$R(p^2/m^2)$ in (\ref{2.16}) denotes the continuum limit of the remainder term after expanding
$\lb\Gamma_{lat}^{(2)}(p)_{\mu\nu}^{ab}\rb_{f,\,1-loop}$ to 2nd order in $p$. It is convergent 
by power-counting and therefore coincides \cite{Reisz(CMP)} with the corresponding (known) 
continuum term:
\be
R(p^2/m^2)&=&\frac{1}{4\pi^2}\int_0^1dx\,x(1-x)\log\Big\lb x(1-x)\frac{p^2}{m^2}+1\Big\rb
\nonumber \\
&=&\frac{1}{24\pi^2}\log(p^2/m^2)
+\frac{1}{4\pi^2}\int_0^1dx\,x(1-x)\log\Big\lb x(1-x)+\frac{m^2}{p^2}\Big\rb
\nonumber \\
&\equiv&\frac{1}{24\pi^2}\log(p^2/m^2)+\wR(m^2/p^2)
\label{2.20}
\ee
Substituting (\ref{2.18}) and (\ref{2.20}) in (\ref{2.16}) gives  
\be
\Big\lb\Gamma_{lat}^{(2)}(p)_{\mu\nu}^{ab}\Big\rb_{f,\,1-loop}
=\delta^{ab}(p^2\delta_{\mu\nu}-p_{\mu}p_{\nu})\Big\lb 
\frac{1}{24\pi^2}\log(a^2p^2)+c_I+\wR(m^2/p^2)\Big\rb
\label{2.21}
\ee
Comparing this to the expression (\ref{2.4}) for $\Gamma_{lat}^{(2)}(p)_{\mu\nu}^{ab}$ with
$\nu_1(p)_{lat}$ given by (\ref{2.8}), we see that
\be
c_{lat}^f=c_I+\wR(0)
\label{2.22}
\ee
In fact $\wR(0)$ is precisely the $\bMS$ constant $c_{\bMS}^f\,$: explicit evaluation of the
integral in (\ref{2.20}) at $m=0$ gives $\wR(0)=-5/72\pi^2$. It follows from (\ref{2.12}) that
\be
l_0^f=-c_I
\label{2.23}
\ee
Thus the issue is to determine the constant $c_I$ appearing in the logarithmically divergent
one-loop lattice integral $I(am)$ in (\ref{2.17})--(\ref{2.18}). In this paper we are going to 
derive a compact general formula for $I(am)$ for general lattice fermions which can then be used
to determine $c_I$.


\section{An initial formula for $I(am)$}

In this section we derive an initial general formula for $I(am)$. It is the starting point for
deriving more explicit formulae in the subsequent sections.

Let $D$ be a general lattice Dirac operator which is translation-invariant and transforms 
covariantly under gauge transformations and rotations of the 4-dimensional
Euclidean hypercubic lattice.
The ansatz for the link variables is
\be
U_{\mu}(x)=e^{aA_{\mu}(x+\frac{1}{2}a\hmu)}
\label{3.1}
\ee
($\hmu$=unit vector in the positive $\mu$-direction). The Fourier-transformed
field $\hA_{\mu}(p)$ is defined via
\be
A_{\mu}(x+\frac{1}{2}a\hmu)=\int_p\,\hA_{\mu}(p)\,e^{ip\cdot(x+\frac{1}{2}a\hmu)}\,.
\label{3.2}
\ee
Here and in the following $\int_p(\cdots)\equiv\int_{-\pi/a}^{\pi/a}\,\frac{d^4p}{(2\pi)^4}\,
(\cdots)\,$. The continuum gauge fields are $A_{\mu}(x)=A_{\mu}^a(x)T^a$ where $\{T^a\}$ are
generators for SU(N) normalized such that $tr(T^aT^b)=-\frac{1}{2}\delta^{ab}$. 
(We take the $T^a$'s to be anti-hermitian, absorbing into them the imaginary unit that 
multiplies $A$ in other notations.)

Expanding the link variables in powers of $A$ gives an expansion of the lattice
Dirac operator,
\be 
D=\sum_{n=0}^{\infty}D_n
\label{3.3}
\ee
Translation-invariance implies that each $D_n$ can be expressed in momentum basis in the form
\be
D_n(k',k)&=&a^{n-1}\int_{p_1,\dots,p_n}\,\delta(p_1+\dots+p_n+k-k')\,
d_n(ak\,|\,ap_1,\dots,ap_n)_{\mu_1\cdots\mu_n} \hA_{\mu_1}(p_1)\cdots\hA_{\mu_n}(p_n) 
\nonumber \\
&&\label{3.4}
\ee
The continuum limit requirements on $D$ are 
\be
\frac{1}{a}\,d_0(ak)&\stackrel{a\to0}{\longrightarrow}&i\gamma_{\nu}k_{\nu}+m \label{3.4a} \\
d_1(ak\,|\,ap)_{\mu}&\stackrel{a\to0}{\longrightarrow}&\gamma_{\mu} \label{3.4b}
\ee
The mass $m$ enters via an additive term $am$ in $d_0(ak)$ (we suppress the $m$-dependence
in the notation).

To derive a general formula for $I(am)$ it is useful to exploit the fact, clear from 
(\ref{2.1}), that the one fermion loop contribution to $\Gamma(A)$ is given by the fermion
determinant:
\be
\Gamma(A)_{f,\,1-loop}=-\log\det D
\label{3.5}
\ee
where, in the present lattice setting, $D$ is coupled to the link variables (\ref{3.1})
of the background gauge field $A$. Decomposing $D$ as 
\be
D=D_0+D_{int}
\label{3.6}
\ee
where $D_0$ is the free field operator and $D_{int}$ the interaction part,
we start from $D=D_0(1+D_0^{-1}D_{int})$ to obtain the expansion
\be
\log\det D-\log\det D_0&=&Tr\,\log(1+D_0^{-1}D_{int}) \nonumber \\
&=&\sum_{l=1}^{\infty}\frac{(-1)^{l-1}}{l}Tr\lb(D_0^{-1}D_{int})^l\rb
\label{3.7}
\ee
From this, using $D_{int}=\sum_{n\ge1}D_n$ with $D_n$ given by (\ref{3.4}), an expansion in
powers of $A$ is obtained. It has the general form
\be
\log\det D-\log\det D_0=\sum_{n\ge1}\int_{p_1,\dots,p_{n-1}}
\Gamma_{\det}^{(n)}(p_1,\dots,p_{n-1})_{\mu_1\cdots\mu_n} 
tr \hA_{\mu_1}(p_1)\cdots\hA_{\mu_n}(p_n)
\label{3.8}
\ee
($p_n=-(p_1+\dots+p_{n-1})$). The ``$n$-point function''
$\Gamma_{\det}^{(n)}(p_1,\dots,p_{n-1})_{\mu_1\cdots\mu_n}$ can be Taylor-expanded in powers of
(the components of) $p_1,\dots,p_{n-1}\,$. Gauge invariance and mass-dimension considerations
imply that the terms in the expansion can be combined to take the form\footnote{In order for
the link variables (\ref{3.1}) to transform in the correct way under lattice gauge 
transformations the transformation of the continuum gauge field $A$ needs to be modified in
the lattice setting \cite{Kawai,Reisz(NPB)}. However, one still finds that the mass-dimension
zero function in (\ref{3.9}) must be $\sim S_{YM}(A)$ -- this can be inferred from
the argument in T. Reisz's proof of renormalizability of Lattice QCD \cite{Reisz(NPB)};
see also \S5 of \cite{Kawai}.}
\be
\log\det D-\log\det D_0=-I(am)S_{YM}(A)+\sum_{r\ge1}\frac{1}{m^r}S_r(A)
\label{3.9}
\ee
up to terms which vanish for $a\to0$, where each $S_r(A)$ is a gauge invariant function with 
mass dimension $r\ge1$. The fact that the coefficient of $S_{YM}(A)$ is $-I(am)$ is inferred
from (\ref{2.16}) and (\ref{3.5}).  

It is clear from (\ref{3.9}) and (\ref{3.5}) that $I(am)$ can be evaluated from either the
2-, 3- or 4-point function in (\ref{3.8}). For the 4-point function the relation is
\be
I(am)=-\Gamma_{\det}^{(4)}(0,0,0)_{\mu\nu\mu\nu}\qquad\quad\mu\ne\nu\ ,\ \ \mbox{no sum}
\label{3.10}
\ee
for any choice of $\mu$ and $\nu$ with $\mu\ne\nu$. Usually $I(am)$ is evaluated from the
2-point function via (\ref{2.17}), but we will see in the following that it is advantageous
to use (\ref{3.10}) instead. To evaluate it we start by collecting the terms containing
4 powers of $A$ in the the expansion of $\log\det D$ in (\ref{3.7}). These are
\be
&&{\textstyle \frac{(-1)^3}{4}}\,Tr\lb(D_0^{-1}D_1)^4\rb
\ +\ {\textstyle \frac{(-1)^2}{3}}\,3Tr\lb(D_0^{-1}D_1)^2D_0^{-1}D_2\rb \nonumber \\
&&\quad+\ {\textstyle \frac{(-1)^1}{2}}\Big(2Tr(D_0^{-1}D_1D_0^{-1}D_3)
\ +\ Tr\lb(D_0^{-1}D_2)^2\rb\Big)\ +\ {\textstyle \frac{(-1)^0}{1}}\,Tr(D_0^{-1}D_4)
\label{3.11}
\ee
Inserting the expression (\ref{3.4}) for each $D_n\,$, and evaluating the traces in momentum
basis, the $\Gamma_{\det}^{(4)}(p_1,p_2,p_3)_{\mu_1\mu_2\mu_3\mu_4}$ in (\ref{3.7}) is
readily worked out. The resulting expression for $I(am)$ from (\ref{3.10}) is as follows.
From the functions $d_n$ in (\ref{3.4}) we define
\be
d_n(k)_{\mu_1\cdots\mu_n}\;\equiv\;d_n(k\,|\,0,\dots,0)_{\mu_1\cdots\mu_n} 
\label{3.12}
\ee 
Then
\be
I(am)=I_{(1,1,1,1)}(am)+I_{(1,1,2)}(am)+I_{(2,2)}(am)+I_{(1,3)}(am)+I_{(0,4)}(am)
\label{3.13}
\ee
where  
\be
&&I_{(1,1,1,1)}(am)=\frac{1}{4}\int_{-\pi}^{\pi}\frac{d^4k}{(2\pi)^4}\,tr\,
d_0(k)^{-1}d_1(k)_{\mu}\,d_0(k)^{-1}d_1(k)_{\nu}\,d_0(k)^{-1}d_1(k)_{\mu}\,d_0(k)^{-1}
d_1(k)_{\nu} \nonumber \\
&&\label{3.14} \\
&&I_{(1,1,2)}(am)=-\int_{-\pi}^{\pi}\frac{d^4k}{(2\pi)^4}\,tr\,
d_0(k)^{-1}d_1(k)_{\mu}\,d_0(k)^{-1}d_1(k)_{\nu}\,d_0(k)^{-1}d_2(k)_{\mu\nu} \label{3.15} \\
&&I_{(2,2)}(am)=\frac{1}{2}\int_{-\pi}^{\pi}\frac{d^4k}{(2\pi)^4}\,tr\,
d_0(k)^{-1}d_2(k)_{\mu\nu}\,d_0(k)^{-1}d_2(k)_{\mu\nu} \label{3.16} \\
&&I_{(1,3)}(am)=\int_{-\pi}^{\pi}\frac{d^4k}{(2\pi)^4}\,tr\,
d_0(k)^{-1}d_1(k)_{\mu}\,d_0(k)^{-1}\,d_3(k)_{\nu\mu\nu} \label{3.17} \\
&&I_{(0,4)}(am)=-\int_{-\pi}^{\pi}\frac{d^4k}{(2\pi)^4}\,tr\,d_0(k)^{-1}d_4(k)_{\mu\nu\mu\nu}
\label{3.18} 
\ee
for any choice of $\mu$ and $\nu$ with $\mu\ne\nu$ (no sum over repeated indices). The traces
are over spinor indices alone.
Each integral corresponds to an $n\!=\!4$ Feynman diagram in Fig.~\ref{vac1} in sect.~I and the 
notation reflects the structure of the corresponding diagram. The above integrals could have been 
derived directly from the Feynman diagrams, with fermion propagator and vertices read off from 
(\ref{3.4}), but it is easier (and equivalent) to use the fermion determinant expansion as we 
have done above.

The expression of $I(am)$ as the sum of the integrals (\ref{3.14})--(\ref{3.18}) is our
initial general formula. It has a number of advantageous features
which we note in the following.

In light of (\ref{3.4a}) it is clear that the integral $I_{(1,1,1,1)}(am)$
diverges logarithmically for $am\to0$ while the other integrals (\ref{3.15})--(\ref{3.18}) 
are all finite in this limit. The divergent integral necessarily \cite{DA(logdiv)} has the 
general form 
\be
I_{(1,1,1,1)}(am)=\frac{1}{24\pi^2}\log(a^2m^2)+\tilde{c}_I
\label{3.19}
\ee
up to terms which vanish for $a\to0$. Then $c_I$ is given by
\be
c_I=\tilde{c}_I+I_{(1,1,2)}(0)+I_{(2,2)}(0)+I_{(1,3)}(0)+I_{(0,4)}(0)
\label{3.20}
\ee
Next we point out that $I_{(1,1,1,1)}(am)$, and hence $\tilde{c}_I$, {\em depend only on the 
free field lattice Dirac operator}. This is because $d_1(k)_{\mu}$ in (\ref{3.14}) is
determined by $d_0(k)\,$: for any lattice fermion formulation we have
\be
d_1(k)_{\mu}=\frac{1}{i}\,\frac{d}{dk_{\mu}}\,d_0(k)\,.
\label{3.21}
\ee
This is derived in Appendix A. 
Therefore, a change of gauging of $D$ (e.g., by adding staples, clover 
term etc.) affects only the finite integrals (\ref{3.15})--(\ref{3.18}).

A further advantage becomes apparent when considering 
the description of $D$ in terms of lattice paths \cite{Has(path),Gatt(path)} (we will discuss the 
path description more explicitly in sect.~V). It is easy to see that the 
vertex function $d_n(k\,|\,p_1,\dots,p_n)_{\mu_1\cdots\mu_n}$ receives contributions only from
lattice paths which contain a lattice link parallel to the $\mu_1$-axis, preceded at some 
point by a link parallel to the $\mu_2$-axis, preceded at some point by a link parallel
to the $\mu_3$-axis, and so on. In particular, if the lattice paths describing $D$ are
straight lines, as is the case for the Wilson-Dirac, naive and staggered operators, then
$d_n(k)_{\mu_1\cdots\mu_n}$ vanishes unless the $\mu_j$'s are all the same. It follows that the 
finite terms (\ref{3.15})--(\ref{3.18}) all vanish in this case (since $\mu\ne\nu$ there).
Thus, for such lattice Dirac operators, $I(am)$ is given entirely by $I_{(1,1,1,1)}(am)$.
If staples ``$\mu\nu\mu$'' are attached to such an operator then (\ref{3.15})--(\ref{3.17})
are non-vanishing while (\ref{3.18}) still vanishes. If a clover term ($\simeq$ closed paths
around plaquettes) is added then all the finite integrals (\ref{3.15})--(\ref{3.18}) are 
potentially non-vanishing. However, as we will see in sect.~V, $d_3(k)_{\nu\mu\nu}$ and 
$d_4(k)_{\mu\nu\mu\nu}$ both vanish in the clover case, so the terms (\ref{3.17})--(\ref{3.18}) 
also vanish there.

These properties make (\ref{3.13})--(\ref{3.20}) useful for evaluating $I(am)$ 
and $c_I$ in practice for specific lattice fermion formulations, especially for improved 
formulations. E.g., when the improved $D$ differs from the original one by a more complicated 
choice of gauging the only new quantities that need to be evaluated are the finite integrals 
(\ref{3.15})--(\ref{3.18}) with $m\!=\!0$ in $d_0(k)$. The quantities $d_2(k)_{\mu\nu}\,$, 
$d_3(k)_{\nu\mu\nu}$ and $d_4(k)_{\mu\nu\mu\nu}$ appearing in their integrands are generally easy 
to determine in practice as we will see in sect.~V. Besides these, the integrands only involve 
the free fermion $d_0(k)$ and its derivative (recall (\ref{3.21})).

Taking the initial formula for $I(am)$ in this section as starting point, we go on to derive 
more explicit expressions in the following two sections.

\section{Formulae for $I_{(1,1,1,1)}(am)$}

For simplicity we use the notations 
$d_0\equiv d_0(k)$ and $\partial_{\mu}\equiv\frac{\partial}{\partial k_{\mu}}$ in the following.
We assume that 
\be
\Delta_0\;\equiv\;d_0^{\dagger}d_0=d_0d_0^{\dagger}
\label{4.1}
\ee
is a scalar, i.e., trivial in spinor space. Note that this is a free field statement; it holds 
for all lattice Dirac operators of current interest (naive, staggered, Wilson, overlap,...).

Recalling (\ref{3.21}): $d_1(k)_{\mu}=-i\partial_{\mu}d_0\,$, and using the relations
\be
d_0^{-1}=\frac{d_0^{\dagger}}{\Delta_0}\qquad,\qquad 
\partial_{\mu}d_0^{-1}=\frac{1}{\Delta_0}\Big(\partial_{\mu}d_0^{\dagger}
-\frac{\partial_{\mu}\Delta_0\,d_0^{\dagger}}{\Delta_0}\Big)\,,
\label{4.2}
\ee
evaluation of (\ref{3.14}) leads to
\be
I_{(1,1,1,1)}(am)=\frac{1}{24}\int_{-\pi}^{\pi}\frac{d^4k}{(2\pi)^4}\,
\frac{tr X_{\mu\nu}(k)}{\Delta_0(k)^2}
\label{4.3}
\ee
for any choice of $\mu,\nu$ with $\mu\ne\nu\,$, where
\be
X_{\mu\nu}&=&
-6d_0^{\dagger}\partial_{\mu}(\partial_{\nu}d_0\,\partial_{\mu}d_0^{\dagger}\,\partial_{\nu}d_0)
+\partial_{\mu}^2(\partial_{\nu}\Delta_0\,d_0^{\dagger}\,\partial_{\nu}d_0)
-\partial_{\mu}^2(\Delta_0\,\partial_{\nu}d_0^{\dagger}\,\partial_{\nu}d_0)
\nonumber \\
&&+\ \partial_{\nu}(\partial_{\mu}^2\Delta_0\,d_0^{\dagger}\,\partial_{\nu}d_0)
-2\partial_{\mu}^2\Delta_0\,\partial_{\nu}d_0^{\dagger}\,\partial_{\nu}d_0
\label{4.4}
\ee
The details of the calculation are given in appendix B. To evaluate this expression further, 
we now assume that $d_0$ (the free field momentum representation of 
the lattice Dirac operator $D$) has the general form
\be
d_0=i\gamma_{\sigma}\rho_{\sigma}+\lambda
\label{4.5}
\ee
where $\rho_{\sigma}(k)$ and $\lambda(k)$ (which includes the mass term) are real scalar
functions, and the gamma-matrices are hermitian, so that 
$d_0^{\dagger}=-i\gamma_{\sigma}\rho_{\sigma}+\lambda$ and $\Delta_0=\rho^2+\lambda^2$.
This is the typical free field form of lattice Dirac operators 
of interest in practice; it covers the naive, staggered, Wilson, and overlap operators
and their improved versions. 
For simplicity we make the further assumption that
\be
\partial_{\mu}\partial_{\nu}d_0=0\quad\quad\mbox{for $\ \mu\ne\nu$}\,.
\label{4.6}
\ee
This holds for Wilson, naive and staggered fermions, also when the Naik term is
included, but does not hold for the overlap operator.
(The extention of the following to the general case $\partial_{\mu}\partial_{\nu}d_0\ne0$ 
is straightforward but tedious, and we defer it to a future article where the 
specific results for the overlap operator will be derived.)

Substituting (\ref{4.5}) into (\ref{4.3})--(\ref{4.4}),
a calculation using (\ref{4.6}) and the other mentioned properties gives 
\be
I_{(1,1,1,1)}(am)=\frac{1}{12}\int_{-\pi}^{\pi}\frac{d^4k}{(2\pi)^4}\,
\frac{Y_{\mu\nu}(k)}{(\rho(k)^2+\lambda(k,am)^2)^2}\qquad\qquad\quad(\mu\ne\nu)
\label{4.7}
\ee
where  
\be
Y_{\mu\nu}&=&
\partial_{\mu}^2\rho_{\mu}^2\,\partial_{\nu}^2\rho_{\nu}^2
-12(\partial_{\mu}\rho_{\mu})^2(\partial_{\nu}\rho_{\nu})^2 
+2\partial_{\mu}^2\lambda^2\,\partial_{\nu}^2\rho_{\nu}^2
-24(\partial_{\mu}\lambda)^2(\partial_{\nu}\rho_{\nu})^2 \nonumber \\
&&-\ 12\partial_{\mu}^2\lambda\,\partial_{\nu}\lambda\,\rho_{\nu}\,\partial_{\nu}\rho_{\nu} 
-4(\partial_{\mu}\lambda)^2\lambda\,\partial_{\nu}^2\lambda
+4\lambda^2\partial_{\mu}^2\lambda\,\partial_{\nu}^2\lambda\,.
\label{4.8}
\ee
From this, using integration by parts (in a similar way to the calculations in appendix B), 
a more compact expresion can be obtained:
\be
I_{(1,1,1,1)}(am)=-\int_{-\pi}^{\pi}\frac{d^4k}{(2\pi)^4}\,\Big(\,
\frac{\lb(\partial_{\mu}\rho_{\mu})^2+(\partial_{\mu}\lambda)^2\rb
\lb(\partial_{\nu}\rho_{\nu})^2+(\partial_{\nu}\lambda)^2\rb}{\Delta_0^2}
-{\textstyle \frac{1}{2}}(\partial_{\mu}\partial_{\nu}\log\Delta_0)^2\Big)
\label{4.8a}
\ee
(recall $\Delta_0=\rho^2+\lambda^2$).
Although this expression is more compact, the preceding one (\ref{4.7})--(\ref{4.8})
seems more useful for evaluating $I_{(1,1,1,1)}(am)$ in practice.

In the remainder of this section we show that these general formulae
readily reproduce the previously known results for the specific cases of (unimproved)
naive, staggered and Wilson fermions. (Recall from sect.~III that $I(am)$ is given entirely
by $I_{(1,1,1,1)}(am)$ in these cases.)


\subsection{Naive and staggered fermions}

For a naive fermion, 
$\rho_{\sigma}=\sin k_{\sigma}\,$, $\lambda=am$ and (\ref{4.8}) reduces to
\be
Y_{\mu\nu}(k)=4\cos2k_{\mu}\cos2k_{\nu}-12\cos^2k_{\mu}\cos^2k_{\nu}  
\label{4.9}
\ee
up to terms which are $O(a)$ for $a\to0$. 
Since a naive fermion decomposes into 4 degenerate staggered fermions \cite{Smit}, 
$I(am)$ for a staggered fermion is obtained by dividing
the naive fermion result by $4$. The invariance of (\ref{4.9}) under 
$k_{\sigma}\to k_{\sigma}+\pi$ allows the integration domain in (\ref{4.7}) to be 
restricted to $[-\pi/2,\pi/2]^4$ at the expense of an overall factor 16. Thus we find
\be
I(am)=N_t\int_{-\pi/2}^{\pi/2}\frac{d^4k}{(2\pi)^4}\;
\frac{\frac{1}{3}\cos2k_{\mu}\cos2k_{\nu}-\cos^2k_{\mu}\cos^2k_{\nu}}
{\Big((am)^2+\sum_{\sigma}\sin^2k_{\sigma}\Big)^2}
\label{4.10}
\ee
up to terms which vanish for $a\to0$, where $N_t$ is the number of fermion ``tastes''
(16 for a naive fermion, 4 for a staggered fermion). 
This is precisely the expression for $I(am)$ derived previously in Eq.~(6.8) of 
\cite{Weisz(NPB)} where the contribution to the gluonic 2-point function from a 
naive/staggered fermion loop was evaluated. ($I(am)$ corresponds to $\Pi^f$ in 
\cite{Weisz(NPB)}.)

After changing variables by $k_{\sigma}\to k_{\sigma}/2$ the integral (\ref{4.10}) can be 
evaluated by the method of Ref.\cite{Burgio}. It expresses the integral in terms
of certain basic lattice integrals that were evaluated numerically to high precision in
\cite{Burgio}. We find the following result:
\be
I(am)=N_t\Big\lb\,\frac{1}{24\pi^2}\Big(\log(2am)^2+\gamma_E-F_0\Big)
+\frac{7}{144}Z_0\Big\rb 
\label{4.12}
\ee
where $\gamma_E$ is the Euler constant and $F_0\,$, $Z_0$ are numerical constants listed to
high numerical accuracy in Table 1 of \cite{Burgio}. Thus the divergent part has the 
correct universal structure, and the constant $c_I$ ($=-l_0^f\,$) that we are after is given by 
\be
-c_I/N_t&=&\frac{1}{24\pi^2}\Big(F_0-\gamma_E-\log(4)\Big)
-\frac{7}{144}Z_0 \nonumber \\
&=&0.0026247621012431485...
\label{4.13}
\ee
In Ref.~\cite{Weisz(NPB)} this constant was denoted $P_4$ and our value agrees with the one in
Eq.~(6.16) of that paper. We have obtained it to much higher numerical precision here 
though, thanks to the high accuracy to which $F_0$ and $Z_0$ were evaluated in \cite{Burgio}.


\subsection{Wilson fermions}

In this case $\rho_{\sigma}=\sin k_{\sigma}\,$, 
$\lambda=r\sum_{\alpha}(1-\cos k_{\alpha})+am$ where $r$ is the Wilson parameter, 
and (\ref{4.8}) reduces to
\be
Y_{\mu\nu}(k)
&=&4\cos2k_{\mu}\cos2k_{\nu}-12\cos^2k_{\mu}\cos^2k_{\nu} \nonumber \\
&&+\ r^2\Big(8\sin^2k_{\mu}(\cos2k_{\nu}-3\cos^2k_{\nu})-6\sin2k_{\mu}\sin k_{\mu}\cos k_{\nu}
+16\cos2k_{\mu}\cos k_{\nu}\sum_{\alpha}\sin^2(k_{\alpha}/2)\Big) \nonumber \\
&&+\ r^4\Big(-8\sin^2k_{\mu}\cos k_{\nu}\sum_{\alpha}\sin^2(k_{\alpha}/2)
+16\cos k_{\mu}\cos k_{\nu}\Big(\sum_{\alpha}\sin^2(k_{\alpha}/2)\Big)^2\,\Big) \nonumber \\
&&\label{4.14}
\ee
up to terms which are $O(a)$ for $a\to0$.
Substituting this into (\ref{4.7}) 
we recover the previous result of Kawai {\em et al.}
for the one fermion loop contribution in Eq.(3.24)--(3.25) of
Ref.~\cite{Kawai}. (Dimensional regularization of the infrared divergence was used there,
but the result is easily transformed into mass-regularized form (cf.~\S5.1 of \cite{Burgio})
and then coincides with ours.) 
The constant $c_I$ is related to the constant $L$ defined there by
\be
c_I=\frac{1}{24\pi^2}(\gamma_E-\log(4\pi))+\frac{1}{2}L
\label{4.16}
\ee
The integral $I(am)$ can again be evaluated by the method of Ref.~\cite{Burgio}; in fact this 
was done in that paper for $r=1$, and the expression for $L$ in terms of the basic constants is 
given in Eq.(6.8) of \cite{Burgio}. 
From that, $l_0^f$ is obtained to high precision in the Wilson fermion 
case:
\be
l_0^f=-c_I=0.00669599933173308...
\label{4.17}
\ee

The Wilson fermion case was also considered independently by P.~Weisz in \cite{Weisz(PLB)}.
Note that for Wilson fermions $(\partial_{\mu}\rho_{\mu})^2+(\partial_{\mu}\lambda)^2
=\sin^2k_{\mu}+\cos^2k_{\mu}=1$,
hence (\ref{4.8a}) simplifies to 
\be
I(am)=-\int_{-\pi}^{\pi}\frac{d^4k}{(2\pi)^4}\,\Big(\,\frac{1}{\Delta_0^2}
-{\textstyle \frac{1}{2}}(\partial_{\mu}\partial_{\nu}\log\Delta_0)^2\Big)
\label{4.18}
\ee
thus reproducing Weisz's expression in Eq.~(13) of \cite{Weisz(PLB)}. 
The constant $l_0^f$, denoted as $P_3$ there, was evaluated numerically but to low precision
-- a systematic error of $\approx 2\%$ was estimated, and this is indeed 
the case when comparing the value for $P_3$ in Eq.(22) of \cite{Weisz(PLB)} with the high 
precision value in (\ref{4.17}) above.

\section{Evaluation of the finite integrals}

In this section we derive general formulae for the finite integrals (\ref{3.15})--(\ref{3.18}) 
and describe how these can be straightforwardly evaluated, using staple and clover terms
as illustrations.

\noindent{\em Evaluation of $I_{(1,1,2)}(am)$.} Starting from (\ref{3.15}), 
calculations of the same type as in appendix B lead to
\be
I_{(1,1,2)}(am)=-\int_{-\pi}^{\pi}\frac{d^4k}{(2\pi)^4}\,
\frac{tr\lb \partial_{\mu}d_0^{\dagger}\,\partial_{\nu}d_0\,d_0^{\dagger}\,d_{2\mu\nu}
-\frac{1}{2}\partial_{\mu}(d_0^{\dagger}\,\partial_{\nu}d_0\,d_0^{\dagger}\,d_{2\mu\nu})\rb}
{\Delta_0^2}
\label{5.1}
\ee
(for any choice of $\mu,\nu$ with $\mu\ne\nu$)
with the notations $d_0=d_0(k)\,$, $\Delta_0=d_0^{\dagger}(k)d_0(k)$ and 
$d_{2\mu\nu}=d_2(k)_{\mu\nu}$.
Specializing as before to the case where $d_0$ has the general form
$d_0=i\gamma_{\sigma}\rho_{\sigma}+\lambda$ and satisfies $\partial_{\mu}\partial_{\nu}d_0=0$
for $\mu\ne\nu$, and {\em assuming} that the gamma-matrix structure of $d_{2\mu\nu}$ has the 
general form
\be
d_2(k)_{\mu\nu}=\gamma_{\mu}\gamma_{\nu}e_{\mu\nu}(k)\quad\quad(\mu\ne\nu)
\label{5.2}
\ee
(\ref{5.1}) leads to
\be
I_{(1,1,2)}(am)=-\int_{-\pi}^{\pi}\frac{d^4k}{(2\pi)^4}\,
\frac{4(\rho_{\mu}\partial_{\nu}\rho_{\nu}\partial_{\mu}\lambda
+\rho_{\nu}\partial_{\mu}\rho_{\mu}\partial_{\nu}\lambda
-\partial_{\mu}\rho_{\mu}\partial_{\nu}\rho_{\nu}\lambda)e_{\mu\nu}}
{(\rho^2+\lambda^2)^2}
\label{5.3}
\ee
(There are no terms involving derivatives of $e_{\mu\nu}(k)$ since these all cancel out.)

\noindent{\em Evaluation of $I_{(2,2)}(am)$.} From (\ref{3.16}) we obtain
\be
I_{(2,2)}(am)=\frac{1}{2}\int_{-\pi}^{\pi}\frac{d^4k}{(2\pi)^4}\,
\frac{tr\,d_0^{\dagger}\,d_{2\mu\nu}\,d_0^{\dagger}\,d_{2\mu\nu}}{\Delta_0^2}
\label{5.4}
\ee
Specializing as before, this leads to
\be
I_{(2,2)}(am)=-2\int_{-\pi}^{\pi}\frac{d^4k}{(2\pi)^4}\,
\frac{\lambda^2\,e_{\mu\nu}^2}{(\rho^2+\lambda^2)^2}
\label{5.5}
\ee

\noindent{\em Evaluation of $I_{(1,3)}(am)$.} From (\ref{3.17}) we find
\be
I_{(1,3)}(am)=\int_{-\pi}^{\pi}\frac{d^4k}{(2\pi)^4}\,
\frac{i\,tr\,d_0^{\dagger}\,\partial_{\mu}d_{3\nu\mu\nu}}{\Delta_0}
\label{5.6}
\ee
with the notation $d_{3\nu\mu\nu}=d_3(k)_{\nu\mu\nu}$.
Specializing as before, and assuming that $d_{3\nu\mu\nu}$ has the general form
\be
d_3(k)_{\nu\mu\nu}=\gamma_{\mu}e_{\nu\mu\nu}-if_{\nu\mu\nu}
\label{5.7}
\ee
(\ref{5.6}) leads to
\be
I_{(1,3)}(am)=4\int_{-\pi}^{\pi}\frac{d^4k}{(2\pi)^4}\,
\frac{\rho_{\mu}\partial_{\mu}e_{\nu\mu\nu}+\lambda\partial_{\mu}f_{\nu\mu\nu}}
{\rho^2+\lambda^2}
\label{5.8}
\ee

\noindent{\em Evaluation of $I_{(0,4)}(am)$.} From (\ref{3.18}) we find
\be
I_{(0,4)}(am)=-\int_{-\pi}^{\pi}\frac{d^4k}{(2\pi)^4}\,
\frac{tr\,d_0^{\dagger}\,d_{4\mu\nu\mu\nu}}{\Delta_0}
\label{5.9}
\ee
with the notation $d_{4\mu\nu\mu\nu}=d_4(k)_{\mu\nu\mu\nu}$.
A specialized formula can be worked out as in the previous cases (we omit the details).
Note that terms in $d_{4\mu\nu\mu\nu}$ involving a product of two or more gamma 
matrices give vanishing contribution to the trace in (\ref{5.9}).

To evaluate the integrals in practice one needs to determine $d_2(k)_{\mu\nu}\,$, 
$d_3(k)_{\nu\mu\nu}$ and $d_4(k)_{\mu\nu\mu\nu}\,$ ($\mu\ne\nu$) for the lattice Dirac operator
$D$. This can be done straightforwardly from the description of $D$ in terms of lattice paths,
as we now describe. In the path description, $D$ is expressed as
\be
D\psi(x)=\frac{1}{a}\sum_{\P}c_{\P}\Gamma_{\P}U_{\P}\lb x,\,x+a\Delta_{\P}\rb\,
\psi(x+a\Delta_{\P})
\label{5.10}
\ee
where the sum is over a collection of translation-equivalence classes $\P$ of oriented lattice 
paths. Each has an associated numerical constant $c_{\P}$ and element $\Gamma_{\P}$ of the 
Clifford algebra generated by the gamma-matrices. 
Associated with each equivalence class $\P$ is a 
vector $\Delta_{\P}$ with integer components: it is the difference in lattice units
betwen the start- and end-points. $U_{\P}\lb x,\,x+a\Delta_{\P}\rb$ denotes the product of the
link variables along the representative path for $\P$ starting at $x+a\Delta_{\P}$ and ending
at $x$. 

The Clifford algebra-valued functions $d_n(k|p_1,\dots,p_n)_{\mu_1\cdots\mu_n}$ in the expansion
(\ref{3.3})--(\ref{3.4}) of $D$ are found by expanding the link variable products
$U_{\P}\lb x,\,x+a\Delta_{\P}\rb$ in (\ref{5.10}) in powers of the continuum gauge field $A$.
In the present case we only need to determine these functions at vanishing ``external momenta'',
i.e., $d_n(k)_{\mu_1\cdots\mu_n}=d_n(k|0,\dots,0)_{\mu_1\cdots\mu_n}$. 
Therefore we can take the continuum gauge fields
$\{A_{\mu}\}$ to be constants. Then $U_{\P}\lb x,\,x+a\Delta_{\P}\rb$ is independent of $x$,
and its expansion in powers of the continuum gauge field has the form
\be
U_{\P}=\sum_n u_{\mu_1\cdots\mu_n}^{\P}\,A_{\mu_1}\cdots A_{\mu_n}
\label{5.11}
\ee
The functions $d_n(k)_{\mu_1\cdots\mu_n}$ are then given by   
\be
d_n(k)_{\mu_1\cdots\mu_n}=\sum_{\P}d_n^{\P}(k)_{\mu_1\cdots\mu_n}
\label{5.12}
\ee
where
\be
d_n^{\P}(k)_{\mu_1\cdots\mu_n}=c_{\P}\Gamma_{\P}u_{\mu_1\cdots\mu_n}^{\P}\,
e^{i\Delta_{\P}\cdot k}
\label{5.13}
\ee

Thus, given the description (\ref{5.10}) of $D$ in terms of lattice paths, the 
problem of determining $d_n(k)_{\mu_1\cdots\mu_n}$ is reduced to determining the coefficients
$u_{\mu_1\cdots\mu_n}^{\P}$ in the expansion (\ref{5.11}) of the link variable product
$U_{\P}$ with constant $A$, which is generally straightforward in practice. 
In the present case things are further simplified since we only need to know 
$d_n(k)_{\mu_1\cdots\mu_n}$ when $\mu_1\cdots\mu_n$ is $\mu\nu\,$, $\nu\mu\nu\,$, and
$\mu\nu\mu\nu$. Since $\mu\ne\nu$ we can replace each link variable $U_{\sigma}^{\pm1}$ appearing
in the product $U_{\P}$ by
\be
U_{\sigma}^{\pm1}\ \to\ \left\{
\begin{array}{ll}
1\pm A_{\sigma} & \ \mbox{if $\ \sigma\in\lbrace\mu,\nu\rbrace$}\\
1 & \ \mbox{if $\ \sigma\notin\lbrace\mu,\nu\rbrace$}\\
\end{array}
\right.
\label{5.14}
\ee
To illustrate this, consider the case of a 
``$\nu\mu\nu$ staple'' $U_{\P}=U_{\nu}U_{\mu}U_{\nu}^{-1}\,$. The relevant parts of the
expansion are found by
\be
U_{\nu}U_{\mu}U_{\nu}^{-1}&\to&(1+A_{\nu})(1+A_{\mu})(1-A_{\nu}) \nonumber \\
&&\ =\ -A_{\mu\nu}-A_{\nu}A_{\mu}A_{\nu}+\mbox{other}
\label{5.15}
\ee
where ``other'' refers to terms which are not proportional to $A_{\mu}A_{\nu}\,$,
$A_{\nu}A_{\mu}A_{\nu}$ or $A_{\mu}A_{\nu}A_{\mu}A_{\nu}$ and therefore do not contribute to
$d_2(k)_{\mu\nu}\,$, $d_3(k)_{nu\mu\nu}$ or $d_4(k)_{\mu\nu\mu\nu}\,$. Thus the relevant
contributions from the staple are 
\be
d_2^{\P}(k)_{\mu\nu}=-c_{\P}\Gamma_{\P}\,e^{ik_{\mu}}\quad,\quad
d_3^{\P}(k)_{\nu\mu\nu}=-c_{\P}\Gamma_{\P}\,e^{ik_{\mu}}\quad,\quad
d_4^{\P}(k)_{\mu\nu\mu\nu}=0
\label{5.16}
\ee
where $c_{\P}\Gamma_{\P}$ are whatever factors that accompany the staple $U_{\P}$ in the
expression (\ref{5.10}) for the lattice Dirac operator.

As another example we now consider the Sheikholeslami-Wohlert clover term for Wilson
clover fermions \cite{clover}:
\be
C=c_{sw}\frac{i}{4}\sum_{\alpha\beta}\sigma_{\alpha\beta}P_{\alpha\beta}
\label{5.17}
\ee
where $c_{sw}$ is a tunable constant, $\sigma_{\alpha\beta}=\frac{i}{2}\lb\gmu\,,\,\gnu\rb$,
and $P_{\alpha\beta}(x)$ is a sum of products of link variables around oriented plaquettes
in the $\alpha\beta$-plane starting and ending at $x$. The relevant contributions in this case
come from the $\mu\nu$ parts:
\be
c_{sw}{\textstyle \frac{i}{4}}(\sigma_{\mu\nu}P_{\mu\nu}+\sigma_{\nu\mu}P_{\nu\mu})
=-c_{sw}{\textstyle \frac{1}{2}}\gmu\gnu P_{\mu\nu}
\label{5.18}
\ee
For constant gauge fields, $P_{\mu\nu}$ is explicitly given by
\be
P_{\mu\nu}=\frac{1}{8}\left\lb
\begin{array}{l}
U_{\mu}U_{\nu}U_{\mu}^{-1}U_{\nu}^{-1}+U_{\nu}U_{\mu}^{-1}U_{\nu}^{-1}U_{\mu} \\
+\ U_{\mu}^{-1}U_{\nu}^{-1}U_{\mu}U_{\nu}+U_{\nu}^{-1}U_{\mu}U_{\nu}U_{\mu}^{-1} \\
-\ (\,\mu\leftrightarrow\nu\,) \\
\end{array}
\right\rb
\label{5.19}
\ee
Expanding this as described above, the relevant expansion coefficients are found to be 
$u_{\mu\nu}=1\,$, $u_{\nu\mu\nu}=0$, and $u_{\mu\nu\mu\nu}=0$. Since the plaquette paths are 
closed we have $\Delta_{\P}=0$ in all cases. Using this, 
it follows from (\ref{5.12})--(\ref{5.13}) that
\be
d_2(k)_{\mu\nu}=-c_{sw}{\textstyle \frac{1}{2}}\gmu\gnu\quad,\quad 
d_3(k)_{\nu\mu\nu}=0\quad,\quad d_4(k)_{\mu\nu\mu\nu}=0
\label{5.20}
\ee
independent of $k$. Note that, as discussed in sect.~III, there are no contributions from the 
Wilson-Dirac operator since in its path description the paths are all straight lines.

The finite integral contributions to $I(am)$ can now be determined in the case of Wilson clover
fermions. By (\ref{5.20}) and the previous formulae, $I_{(1,3)}=I_{(0,4)}=0$. The other integrals
are determined by substituting $\rho_{\mu}=\sin k_{\mu}\,$,  
$\lambda=r\sum_{\sigma}(1-\cos k_{\sigma})+am$ and $e_{\mu\nu}=-c_{sw}/2$ into (\ref{5.3}) and
(\ref{5.5}). Taking the Wilson parameter to be $r=1$ and evaluating the integrals by the
method of Ref.~\cite{Burgio} we find, in the $a\to0$ limit,
\be
I_{(1,1,2)}(0)/c_{sw}&=&\frac{2}{3}\F(1,0)-\F(2,-1) \nonumber \\
&=&0.005046714024535753066...
\label{5.21}
\ee
and
\be
I_{(2,2)}(0)/c_{sw}^2&=&-\frac{1}{8}\F(2,-2) \nonumber \\
&=&-0.02984346719542684815...
\label{5.22}
\ee
where $\F(1,0)\,$, $\F(2,-1)$ and $\F(2,-2)$ are certain basic convergent one loop lattice 
integrals defined in \S4 of Ref.~\cite{Burgio}, whose numerical values are given to high 
precision in Table 2 of that paper.

Recalling (\ref{3.20}), collecting the results for Wilson clover fermions with $r=1$ we have 
\be
l_0^f=-c_I=\mbox{(\ref{4.17})}-c_{sw}\mbox{(\ref{5.21})}-c_{sw}^2\mbox{(\ref{5.22})}
\label{5.23}
\ee
((\ref{4.17}) means the numerical constant given in Eq.~(\ref{4.17}), etc.)
This agrees with the previous literature but gives the numerical constants to much greater
precision. The previously most precise values were those in Eq.~(14) of Ref.~\cite{Bode-Pan} 
where (\ref{5.21}) and (\ref{5.22}) were given to 11 decimal places. We have obtained them 
here to 20 decimal places, thanks to the high precision with which the basic integrals in
Ref.~\cite{Burgio} were evaluated. For earlier results for these quantities, 
see \cite{Bode-Weisz} and the references therein.

\section{Concluding remarks}

The focus in this paper has been on deriving the general integral formulae for $I(am)$, 
confirming their correctness by checking that they reproduce the known results in the cases
of staggered, Wilson and clover fermions, and in the process developing general techniques for
evaluating the formulae. In doing this we were able to express
$I(am)$ in those cases in terms of basic one loop 
lattice integrals that have already been evaluated to high precision in \cite{Burgio}. 
This had already been
done in \cite{Burgio} in the Wilson fermion case, but the results for the staggered and
clover cases ((\ref{4.13}) and (\ref{5.21})--(\ref{5.22}), respectively) are presented here
for the first time. The further applications of the general formulae are left for future 
work, and in the following we discuss some possibilities for this.

Obvious targets for future applications are the various improved versions of staggered
fermions. These formulations involve ``smearing'' of the link variables to reduce flavor 
symmetry violations; specifically there is the ``Fat-7'' link \cite{Fat7} and HYP link 
\cite{HYP}. Since these differ from unimproved staggered fermions only by a choice of gauging,
evaluating $I(am)$ should be straighforward: expand the relevant products 
of link variables in powers of the constant continuum gauge field to determine 
$d_2(k)_{\mu\nu}\,$, $d_3(k)_{\nu\mu\nu}$ and $d_4(k)_{\mu\nu\mu\nu}$ as described in sect.~V;
then, from the formulae in sect.~V the convergent integrals $I_{(1,1,2)}$, $I_{(2,2)}$, 
$I_{(1,3)}$ and $I_{(0,4)}$ can be explicitly evaluated by the method of Ref.~\cite{Burgio};
this will express the integrals in terms of basic lattice integrals that have already been
calculated to high precision. Adding these to the already known $I_{(1,1,1,1)}(am)$ 
($=I(am)$ for unimproved staggered fermions) then gives $I(am)$ in these improved cases. 

Of more interest, however, is the case of $O(a^2)$
improved ``Asqtad'' staggered fermions \cite{asqtad} that are currently used by the MILC
collaboration to generate the ensembles used in high precision lattice simulations 
\cite{Davies(PRL)}. Besides smearing of link variables, this formulation also contains the
Naik term \cite{Naik} which modifies the free field staggered Dirac operator. Therefore, 
$I_{(1,1,1,1)}$ is not the same as for unimproved staggered fermions in this case, and
the method and results of Ref.~\cite{Burgio} do not apply. $I(am)$ may still
be readily determined in this case from the general formulae and techniques of this paper, but
it will be necessary to numerically evaluate the one loop lattice integrals that arise.
(The approach of Ref.~\cite{Becher} could be used for this.) 
The relation between the $\bMS$ coupling and bare lattice coupling
has already been determined to 2 loops for Asqtad staggered fermions via symbolic computer
program \cite{Trottier(review)}. Determining the fermionic contribution to the one loop 
coefficient independently via the (semi-)analytic approach of the present paper would provide 
a useful check on the computer program.

Another future application is to overlap fermions. The fermionic contribution to the one loop
relation between the $\bMS$ coupling and bare lattice coupling in this case was calculated via
symbolic computer program in \cite{Vicari}, and the two loop relation was 
subsequently calculated in \cite{Const-Pan1,Const-Pan2}. Application of the results of the 
present paper will allow the one fermion loop contribution to be obtained from numerical
evaluation of one loop lattice integrals without the need for symbolic computer programs.
Reproducing the result of \cite{Vicari} in this way will be a useful check on the computer 
program, which the two loop result \cite{Const-Pan1,Const-Pan2} also relies on. It will
also allow the one fermion loop contribution to be calculated more easily and to higher precision
for any value of the overlap parameter that one wishes to consider. I emphasize that, in the 
approach of the present paper, the technical problem of calculating the one fermion loop 
contribution for overlap fermions is greatly reduced compared to the usual approach followed in 
\cite{Vicari}. The simplification comes about because the determination of the
quantities $d_2(k)_{\mu\nu}\,$, $d_3(k)_{\nu\mu\nu}$ and $d_4(k)_{\mu\nu\mu\nu}$ in the 
expansion of the overlap Dirac operator can be done with constant continuum gauge fields.
Expansion of the overlap Dirac operator in powers of constant continuum gauge fields is 
relatively straightforward, and has already been successfully used \cite{DA(overlap)}
to reproduce results obtained via symbolic computer program in another context
\cite{Horvath(overlap)}.\footnote{Ref.~\cite{DA(overlap)} is the work mentioned in 
the concluding section of \cite{Horvath(overlap)}.}

Having found useful general formulae for the fermionic contribution to the coefficient
in the one loop relation between the $\bMS$ and bare lattice couplings, a natural question
is whether a similar approach is possible for the contribution from the gauge
and ghost loops. In fact this seems quite possible. In the background field approach the 
gauge field in the action is $A+gQ$, and it is the part quadratic in the quantum fluctuation 
fields $Q$ that determines the one loop contribution to the effective action. Coming from the
functional integral of a quadratic term, it is clear that this can be expressed
as a functional determinant, both in the continuum and lattice
settings. There is also the Faddeev-Popov determinant through which the ghosts arise.
One can then expand the determinants in powers of $A$ as done for the fermion determinant
in this paper. The constant to be determined in this case, $c_{lat}^g$ in (\ref{2.8}), can again
be found from the quartic term in the expansion (via Ward Indentities). In the fermionic 
case considered in the present paper, the possibility to introduce an infrared regulator mass 
was crucial for deriving the general formulae. This can also be done in the 
gluonic--ghost case: note that $Q$ transforms under gauge transformations by $Q\to UQU^{-1}$ 
so a regulator mass term $m\,\mbox{tr}\,Q^2$ may be introduced without breaking gauge invariance. 
This is currently being
pursued, and I hope to be able to present general formulae for the gluonic--ghost contribution
to the one loop coefficient in future work. The hope is that this may allow allow a 
(semi-)analytic calculation of the one loop contribution in the case of improved lattice 
gauge actions, which has recently been evaluated via symbolic computer program in
\cite{Skour-Pan}.

Finally, the results of this paper are relevant for a previous proposal for
constructing the gauge field action on the lattice from the lattice fermion determinant
\cite{Horvath}. Set $\eta=am$ and regard $m=\eta/a$ as
a function of $\eta$ and $a$. Expanding 
$\Gamma_{\det}^{(n)}(p_1,\dots,p_{n-1})_{\mu_1\cdots\mu_n}$ in (\ref{3.8}) in powers of the
momenta, without taking the $a\to0$ limit, leads on dimensional grounds to the
following variant of (\ref{3.9}):
\be
\log\det D(\eta)-\log\det D_0(\eta)=-I(\eta)S_{YM}(A)+\sum_{r\ge1}a^r\widetilde{S}_r(A,\eta)
\label{6.1}
\ee
where $\widetilde{S}_r(A,\eta)$, a function of $A$ and $\eta$, has mass-dimension $r$.
Now taking $a\to0$ we obtain\footnote{Since $m=\eta/a$, taking $a\to0$ with $\eta$ fixed
amounts to taking a simultaneous continuum and large mass limit. The fact that the large mass 
limit of the lattice fermion determinant effectively gives a lattice gauge action was discussed 
before in \cite{DeGrand-H}.}   
\be
\lim_{a\to0}\ \Big(\log\det D(\eta)-\log\det D_0(\eta)\Big)=-I(\eta)S_{YM}(A)
\label{6.2}
\ee
(Strictly speaking this depends on the sum in (\ref{6.1}) being convergent, which requires
the continuum gauge field $A$ to be sufficiently weak.)
Thus the coefficient of the YM gauge action obtained from the lattice fermion determinant
is seen to be $-I(\eta)$, which can be evaluated from the general formulae in this paper after
replacing $am$ by $\eta$.\footnote{The relationship between $I(\eta)$ and the coefficient 
$c^S(\eta)$ in \cite{Horvath} is $I(\eta)=2c^S(\eta)$.}
In the $\eta\to0$ limit we have
\be
I(\eta)\;\stackrel{\eta\to0}{\approx}\;\frac{1}{24\pi^2}\log(\eta^2)+c_I
\label{6.3}
\ee
Although calculation of the constant $c_I$ was the main focus in this paper, the general
formulae can just as well be used to determine $I(\eta)$ for general values of $\eta$. 
The integrals will need to be evaluated numerically for the specific values of $\eta$ that 
one considers though.

\begin{acknowledgments}

I am very grateful to Prof.~Peter Weisz for his constructive suggestions on
the previous (and completely different) version of this paper. At S.N.U. the author is 
supported by the BK21 program. At F.I.U. the author was supported under NSF grant
PHY-0400402. Some of this work was done during the KITP program ``Modern Challenges for Lattice 
Field Theory'' where the author was supported under NSF grant PHY99-0794.

\end{acknowledgments}

\appendix

\section{Derivation of the formula (\ref{3.21}) for $d_1(k)_{\mu}$}

We derive this from the path description (\ref{5.10})--(\ref{5.13}) of the lattice Dirac
operator. From (\ref{5.13}) we have 
\be
d_0^{\P}(k)&=&c_{\P}\,\Gamma_{\P}\,e^{i\Delta_{\P}\cdot k} \label{A1} \\
d_1^{\P}(k)_{\mu}&=&c_{\P}\,\Gamma_{\P}\,u_{\mu}^{\P}\,e^{i\Delta_{\P}\cdot k} \label{A2}
\ee
Thus to show the relation (\ref{3.21}): $d_1(k)_{\mu}=-i\pmu d_0(k)$ it suffices to show
\be
u_{\mu}^{\P}=(\Delta_{\P})_{\mu}
\label{A3}
\ee
From the definition (\ref{5.11}) it is easy to see that $u_{\mu}^{\P}$ counts the number of
links of the path $\P$ that lie along the $\mu$-direction, counted with sign $\pm$ depending on 
whether they are oriented in the positive or negative $\mu$-direction. But this is precisely 
the $\mu$-coordinate of the difference (in lattice units) between the start- and end points of 
(a representative path for) $\P$, i.e., the $\mu$-coordinate of $\Delta_{\P}$. Thus we have
found that (\ref{A3}) holds.

\section{Derivation of the general formula for $I_{(1,1,1,1)}(am)$}

The general formula (\ref{4.3})--(\ref{4.4}) is obtained by evaluating the integrand in the
initial expression (\ref{3.14}) as follows. For notational simplicity we omit the ``$tr$'',
write $d$, $\Delta$ for $d_0$, $\Delta_0$, and use ``$\simeq$'' to denote equality up to
terms which vanish upon taking the trace or terms which are total derivatives and therefore
give vanishing contribution to the integral.
\be
&& d^{-1}\pmu d\,d^{-1}\pnu d\,d^{-1}\pmu d\,d^{-1}\pnu d \nonumber \\
&&\quad=\ \pmu d^{-1}\pnu d\,\pmu d^{-1}\pnu d \nonumber \\
&&\quad=\ \frac{1}{\Delta^2}\Big(\pmu d^{\dagger}-\frac{\pmu\Delta\,d^{\dagger}}{\Delta}\Big)
\pnu d\Big(\pmu d^{\dagger}-\frac{\pmu\Delta\,d^{\dagger}}{\Delta}\Big)\pnu d \nonumber \\
&&\quad\simeq\ \frac{1}{\Delta^2}\pmu d^{\dagger}\pnu d\,\pmu d^{\dagger}\,\pnu d
+\pmu\Big(\frac{1}{\Delta^2}\Big)d^{\dagger}\,\pnu d\,\pmu d^{\dagger}\,\pnu d
-\pmu\Big(\frac{1}{3\Delta^3}\Big)\pmu\Delta\,d^{\dagger}\,\pnu d\,d^{\dagger}\,\pnu d
\nonumber \\
&&\quad\simeq\ -\frac{1}{\Delta^2}\,d^{\dagger}\,\pmu(\pnu d\,\pmu d^{\dagger}\,\pnu d)
+\frac{1}{3\Delta^3}\Big(\pmu^2\Delta\,d^{\dagger}\,\pnu d\,d^{\dagger}\,\pnu d
+\pmu\Delta\,\pmu(d^{\dagger}\,\pnu d\,d^{\dagger}\,\pnu d)\Big) \nonumber \\
&&\label{B.1}
\ee
The second term here is re-expressed using
\be
d^{\dagger}\,\pnu d\,d^{\dagger}\,\pnu d
=(\pnu\Delta-\pnu d^{\dagger}\,d)d^{\dagger}\,\pnu d
=\pnu\Delta\,d^{\dagger}\,\pnu d-\Delta\,\pnu d^{\dagger}\,\pnu d
\label{B.2}
\ee
to get
\be
\frac{1}{3}\pnu\Big(\frac{-1}{2\Delta^2}\Big)\pmu^2\Delta\,d^{\dagger}\,\pnu d
-\frac{1}{3\Delta^2}\pmu^2\Delta\,\pnu d^{\dagger}\,\pnu d
&\simeq&\frac{1}{3\Delta^2}\Big(\frac{1}{2}\pnu(\pmu^2\Delta\,d^{\dagger}\,\pnu d)
-\pmu^2\Delta\,\pnu d^{\dagger}\,\pnu d\Big) \nonumber \\
&&\label{B.3}
\ee  
The third term in (\ref{B.1}) is re-expressed as
\be
\frac{1}{3}\pmu\Big(\frac{-1}{2\Delta^2}\Big)\pmu(d^{\dagger}\,\pnu d\,d^{\dagger}\,\pnu d)
\ \simeq\ \frac{1}{6\Delta^2}\,\pmu^2(d^{\dagger}\,\pnu d\,d^{\dagger}\,\pnu d)
\label{B.4}
\ee
The resulting expression for the integrand, given by the first term in (\ref{B.1}) plus
(\ref{B.3}) plus (\ref{B.4}), leads to the claimed formula (\ref{4.3})--(\ref{4.4}).

\end{document}